# Photonic and plasmonic transition radiation from graphene

Jialin Chen, Hongsheng Chen, and Xiao Lin[*]

*Interdisciplinary Center for Quantum Information, State Key Laboratory of Modern Optical Instrumentation, ZJU-Hangzhou Global Science and Technology Innovation Center, College of Information Science and Electronic Engineering, Zhejiang University, Hangzhou 310027, China.*

[*]*To whom correspondence should be addressed:*
*E-mail: xiaolinbnwj@gmail.com (X. Lin)*

**In the framework of full Maxwell equations, we systematically study the electromagnetic radiation, namely the transition radiation, when a swift electron perpendicularly crosses a monolayer graphene. Based on the plane wave expansion and the Sommerfeld integration, we demonstrate the spatial distribution of this free-electron radiation process in the frequency domain, which clearly shows the broadband excitation of both the photons and graphene plasmons. Moreover, the radiation spectra for the excited photons and graphene plasmons are analytically derived. We find that the excitation of photons and graphene plasmons favours different velocities of the swift electron. To be specific, a higher electron velocity can give rise to the excitation of photons with better directivity and higher intensity, while a lower electron velocity can enable the efficient excitation of graphene plasmons in a broader frequency range. Our work indicates that the interaction between swift charged particles and ultrathin 2D materials or metasurfaces is promising for the design of terahertz on-chip radiation sources.**

**Introduction**

Transition radiation, as a typical free-electron radiation process, was first theoretically proposed by V. L. Ginzburg and I. Frank in 1945 and later experimentally observed in 1959 [1, 2]. Transition radiation would occur, once a charged particle moves across a spatially inhomogeneous region, such as an interface. Remarkably, the occurrence of transition radiation has no specific requirements on the particle velocity $v$. This is drastically different from another well-known phenomenon of free-electron radiation, namely Cherenkov radiation [3-5], whose emergence requires the particle velocity larger than the Cherenkov threshold (i.e., the phase velocity of light in the host material) [6, 7]. Furthermore, the usage of periodic photonic nanostructures or periodic electron bunches can lead to the formation of coherent or resonant transition radiation [8-11], which would further enhance the radiation intensity and directivity. As a result, transition radiation has wide applications for the design of radiation sources, ranging from microwave, THz to X-ray regimes [12-15]. On the other hand, for highly relativistic charged particles, the total energy of photons emitted in transition radiation per interface is proportional to $\gamma$ [15], where $\gamma = \frac{1}{\sqrt{1-v^2/c^2}}$ is the Lorentz factor and $c$ is the speed of light in free space. Such a unique feature makes transition radiation attractive as an underlying mechanism for high-energy particle detectors, namely the transition radiation detector [16-18]. The transition radiation detector is useful for the identification of particles at extremely high energies where other existing technologies such as Cherenkov detectors [19, 20] are not accessible. In addition, since the electron bunches can be focused to a nano-spot, the interaction between fast electrons and 2D materials with the precise spatial control has been widely exploited for the exploration of many emerging properties of 2D materials [21-28], for example, through the electron energy loss spectroscopy (EELS). In turn, due to the active tunability of their optical properties, these 2D materials [29-45], such as graphene, show the promise to provide new degrees of freedom for the flexible control of transition radiation, which is worthy of further studies.

Here, we systematically study the transition radiation from a monolayer graphene, by extending Ginzburg and Frank's theory of transition radiation to graphene physics in the framework of classic electromagnetic wave theory. We theoretically investigate in the frequency domain the spatial distribution and energy spectrum of graphene plasmons excited by a swift electron perpendicularly impacting a monolayer graphene. The similar analysis is also performed on the associated light emission. We find that both graphene plasmons and photons can be efficiently excited during the process of transition radiation. As expected, the more efficient excitation of photons requires a higher electron velocity. Counterintuitively, the more efficient excitation of graphene plasmons prefers a lower electron velocity; such a feature is desirable for many on-chip optoelectronic applications, especially for the design of novel on-chip radiation sources with compacted dimensions.

**Analytical calculation of the radiation field from the transition radiation process**

Figure 1 shows the structural schematic. Here we consider a monolayer graphene parallel to the $xy$ plane and located at the interface between medium 1 ($z < 0$) and medium 2 ($z > 0$). To explore the influence of graphene on the transition radiation, we set both nonmagnetic media to be free space. That is, their relative permittivities and permeabilities are $\varepsilon_{1r} = \varepsilon_{2r} = \varepsilon_r = 1$ and $\mu_{1r} = \mu_{2r} = \mu_r = 1$, respectively.

When the charged particle has a charge $q$ and moves at a velocity $\bar{v} = \hat{z}v$, the induced current density is

$$\bar{J}^q(\bar{r},t) = \hat{z}qv\delta(x)\delta(y)\delta(z - vt) \tag{1}$$

By applying the Fourier transformation or the plane wave expansion [1, 2, 46, 47], the current density and the related electric and magnetic fields can be expressed as,

$$\bar{J}^q(\bar{r},t) = \int \hat{z} j^q_{\omega,\bar{\kappa}_\perp}(z) e^{i(\bar{\kappa}_\perp \cdot \bar{r}_\perp - \omega t)} d\bar{\kappa}_\perp d\omega \tag{2}$$

$$\bar{E}(\bar{r},t) = \int \bar{E}_{\omega,\bar{\kappa}_\perp}(z) e^{i(\bar{\kappa}_\perp \cdot \bar{r}_\perp - \omega t)} d\bar{\kappa}_\perp d\omega \tag{3}$$

$$\bar{H}(\bar{r},t) = \int \bar{H}_{\omega,\bar{\kappa}_\perp}(z) e^{i(\bar{\kappa}_\perp \cdot \bar{r}_\perp - \omega t)} d\bar{\kappa}_\perp d\omega \tag{4}$$

where $\bar{\kappa}_\perp = \hat{x}\kappa_x + \hat{y}\kappa_y$ is the component of wavevector perpendicular to the charge trajectory. From equations (1-2), we readily get $j^q_{\omega,\bar{\kappa}_\perp}(z) = \frac{q}{(2\pi)^3} e^{i\frac{\omega}{v}z}$. Moreover, from the Maxwell equations in the frequency domain, we can further obtain the following equation for $E_{z,\omega,\bar{\kappa}_\perp}$ (i.e., the z component of the electric field $\bar{E}_{\omega,\bar{\kappa}_\perp}$),

$$\frac{\partial^2}{\partial z^2}\left(\varepsilon_r E_{z,\omega,\bar{\kappa}_\perp}\right) + \varepsilon_r \left(\frac{\omega^2}{c^2}\varepsilon_r - \kappa_\perp^2\right) E_{z,\omega,\bar{\kappa}_\perp} = -\frac{i\omega\mu_0 q}{(2\pi)^3}\left(\varepsilon_r - \frac{c^2}{v^2}\right) e^{i\frac{\omega}{v}z} \tag{5}$$

To solve equation (5), we may express $E_{z,\omega,\bar{\kappa}_\perp}$ in each medium as a sum of the charge field $\bar{E}^q_{z,\omega,\bar{\kappa}_\perp}$ and the radiation field $\bar{E}^R_{z,\omega,\bar{\kappa}_\perp}$ [1, 2]. If we solve equation (5) in a homogenous medium, we can directly obtain the charge field in each medium, that is,

$$E^q_{z,\omega,\bar{\kappa}_\perp} = \frac{-iq}{\omega\varepsilon_0(2\pi)^3} \frac{1-\frac{c^2}{v^2\varepsilon_r}}{\varepsilon_r - \frac{c^2}{v^2} - \frac{\kappa_\perp^2 c^2}{\omega^2}} e^{i\frac{\omega}{v}z} \tag{6}$$

The charge field in equation (1) acts equivalently as the incident waves, while the radiation field can be understood as the scattered waves. Due to the momentum conservation, both the incident and scattered waves should have a same component of wavevector parallel to the interface (i.e., $\bar{\kappa}_\perp$). Then without loss of generality, we may further assume the radiation field in each medium with the following form

$$E^R_{z,\omega,\bar{\kappa}_\perp} = \frac{iq}{\omega\varepsilon_0(2\pi)^3} \cdot a \cdot e^{\pm i\frac{\omega}{c}\sqrt{\varepsilon_r - \frac{\kappa_\perp^2 c^2}{\omega^2}}z} \tag{7}$$

Since the radiation field propagates away from the boundary, the $+$ sign should be used for the forward radiation in medium 2, and the $-$ sign should be adopted for the backward radiation in medium 1.

The unknown parameter $a$ in the radiation fields can be solved by enforcing the electromagnetic boundary conditions for the total fields, that is,

$$\hat{n} \times (\bar{E}_{1\perp} - \bar{E}_{2\perp})|_{z=0} = 0, \quad \hat{n} \times (\bar{H}_{1\perp} - \bar{H}_{2\perp})|_{z=0} = \bar{J}_s = \sigma_s \bar{E}_{1\perp}|_{z=0} \tag{8}$$

In the above, $\hat{n} = -\hat{z}$, $\bar{J}_s$ is the induced surface current density due to the existence of the monolayer graphene at the interface, and the subscript $\perp$ stands for the field component parallel to the interface. Here the monolayer graphene is modelled by a surface conductivity $\sigma_s$, which can be readily calculated from the Kubo formula [31]. Without specific specification, we set the chemical potential of graphene to be $\mu_c = 0.4$ eV and the relaxation time to be $\tau = 0.5$ ps in this work [31].

From these boundary conditions, we obtain the amplitude $a_1$ for the backward radiation in medium 1 and the amplitude $a_2$ for the forward radiation in medium 2 as follows

$$a_1 = -a_2 = a = \frac{\frac{v}{c}\frac{\kappa_\perp^2 c^2}{\omega^2 \varepsilon_r} \cdot \frac{\sigma_s}{c\varepsilon_0}}{(1-\frac{v^2}{c^2}\varepsilon_r + \frac{\kappa_\perp^2 v^2}{\omega^2})(2\varepsilon_r + \frac{\sigma_s}{c\varepsilon_0}\sqrt{\varepsilon_r - \frac{\kappa_\perp^2 c^2}{\omega^2}})} \tag{9}$$

Note that the above radiation fields are solved in the Cartesian coordinates. Actually due to the rotational symmetry of our studied structure, all fields can be expressed in the cylindrical coordinates. After some calculations, we have,

$$\bar{E}_m(\bar{r},t) = \bar{E}_m^q(\bar{r},t) + \bar{E}_m^R(\bar{r},t), \bar{H}_m(\bar{r},t) = \bar{H}_m^q(\bar{r},t) + \bar{H}_m^R(\bar{r},t) \tag{10}$$

$$\bar{E}_m^q(\bar{r},t) = \hat{z} \int_{-\infty}^{+\infty} d\omega \frac{-q}{8\pi\omega\varepsilon_0\varepsilon_{mr}} \left(\frac{\omega^2}{c^2}\varepsilon_{mr} - \frac{\omega^2}{v^2}\right) H_0^{(1)}\left(\rho\sqrt{\frac{\omega^2}{c^2}\varepsilon_{mr} - \frac{\omega^2}{v^2}}\right) e^{i\left(\frac{\omega}{v}z - \omega t\right)} +$$

$$\hat{\rho} \int_{-\infty}^{+\infty} d\omega \frac{-q}{8\pi\omega\varepsilon_0\varepsilon_{mr}} (i\frac{\omega}{v})(-\sqrt{\frac{\omega^2}{c^2}\varepsilon_{mr} - \frac{\omega^2}{v^2}} H_1^{(1)}\left(\rho\sqrt{\frac{\omega^2}{c^2}\varepsilon_{mr} - \frac{\omega^2}{v^2}}\right) e^{i\left(\frac{\omega}{v}z - \omega t\right)}) \tag{11}$$

$$\bar{H}_m^q(\bar{r},t) = \hat{\phi} \int_{-\infty}^{+\infty} d\omega \frac{iq}{8\pi} \sqrt{\frac{\omega^2}{c^2}\varepsilon_{mr} - \frac{\omega^2}{v^2}} H_1^{(1)}(\sqrt{\frac{\omega^2}{c^2}\varepsilon_{mr} - \frac{\omega^2}{v^2}}\rho) e^{i\left(\frac{\omega}{v}z - \omega t\right)} \tag{12}$$

$$\bar{E}_m^R(\bar{r},t) = \hat{z} \int_{-\infty}^{+\infty} d\omega \int_0^{+\infty} d\kappa_\perp \cdot \frac{iq}{\omega\varepsilon_0(2\pi)^3} a_m \kappa_\perp (2\pi J_0(\kappa_\perp \rho)) e^{i[(-1)^m\left(\frac{\omega}{c}\sqrt{\varepsilon_{mr} - \frac{\kappa_\perp^2 c^2}{\omega^2}}\right)z - \omega t]} +$$

$$\hat{\rho} \int_{-\infty}^{+\infty} d\omega \int_0^{+\infty} d\kappa_\perp \cdot \frac{iq}{\omega\varepsilon_0(2\pi)^3} a_m((-1)^{m+1}\frac{\omega}{c}\sqrt{\varepsilon_{mr} - \frac{\kappa_\perp^2 c^2}{\omega^2}})(i2\pi J_1(\kappa_\perp \rho)) e^{i[(-1)^m\left(\frac{\omega}{c}\sqrt{\varepsilon_{mr} - \frac{\kappa_\perp^2 c^2}{\omega^2}}\right)z - \omega t]}$$

$$\tag{13}$$

$$\bar{H}_m^R(\bar{r},t) = \hat{\phi} \int_{-\infty}^{+\infty} d\omega \int_0^{+\infty} d\kappa_\perp \frac{iq}{\omega\varepsilon_0(2\pi)^3} a_m(-\omega\varepsilon_0\varepsilon_{mr})(i2\pi J_1(\kappa_\perp \rho)) e^{i[(-1)^m\left(\frac{\omega}{c}\sqrt{\varepsilon_{mr} - \frac{\kappa_\perp^2 c^2}{\omega^2}}\right)z - \omega t]} \tag{14}$$

In the above, we set the subscript $m = 1$ or $2$, and $\kappa_\perp = |\bar{\kappa}_\perp|$.

Based on equations (10-14), all fields created during the process of transition radiation can be numerically obtained. Especially, the excitation of graphene plasmons during the transition radiation process can be identified as the poles in the denominator of $a_m$ in equations (13-14). To be specific, once the denominator of $a_m$ is equal to zero, we can obtain the solution for the dispersion relation of graphene plasmons, that is, $2\varepsilon_r + \frac{\sigma_s}{c\varepsilon_0}\sqrt{\varepsilon_r - \frac{\kappa_\perp^2 c^2}{\omega^2}} = 0$ for the case of $\varepsilon_{1r} = \varepsilon_{2r} = \varepsilon_r$. This way, if the whole system is lossless, we have $\kappa_\perp = +\sqrt{\frac{\omega^2}{c^2}[\varepsilon_r - \left(\frac{2\varepsilon_r\varepsilon_0 c}{\sigma_s}\right)^2]}$ as a real number, and there will be a polariton pole singularity on the real $\kappa_\perp$ axis, rendering the integral in equation (13-14) not well defined. To overcome this ambiguity, the material loss should be assumed for the monolayer graphene (i.e., with a finite value of relaxation time). Then the polariton pole singularity is now located off the real $\kappa_\perp$ axis, and the integral in equation (13-14) becomes unambiguous.

Mathematically, the numerical integration of the radiation fields in equation (13-14) can be readily carried out through the application of the Sommerfeld integration path [48, 49].

**Near-field analysis of the transition radiation from graphene**

Figure 2 shows the radiation field in the region close to the monolayer graphene. For conceptual demonstration, we set $v = 0.8c$, namely the kinematic energy of electrons is 340 keV, which can be achieved in modern electron microscope systems [5]. To investigate the excitation of graphene plasmons, we plot in Fig. 2a the radiation field along a specific line, which is parallel to the graphene plane and only has a vertical distance of 10 nm to the graphene plane. From Fig. 2a, we can clearly see the broadband excitation of graphene plasmons. Meanwhile, as the frequency increases, the wavelength of graphene plasmons decreases in a rapid way. Such a feature can also be seen in Fig. 2b-e, where we plot the near-field radiation field at four different THz frequencies. Especially, at a higher frequency [e.g., 4 THz in Fig. 2e], the wavelength of graphene plasmons is much smaller than that of the excited photons. From Fig. 2b-e, the spatial confinement of graphene plasmons rapidly decreases if the frequency increases. Moreover, Fig. 2b-e show that both the graphene plasmons and photons are excited during the transition radiation process. While the excited photons can freely propagate to the far field, the excited graphene plasmons will decay during the propagation, due to the finite loss of graphene.

The graphene plasmons excited during the transition radiation process can also be analyzed in a quantitative way. That is, we can approximately calculate the spectrum for the excited graphene plasmons. To facilitate the analytical calculation, all material losses are neglected in this part. Under this condition, the energy of excited graphene plasmons is equivalent to the energy of the fields carried by the graphene plasmons at the time of $t \to \infty$. In other words, we can express the electromagnetic energy of the excited graphene plasmons as

$$W^{\mathrm{s}} = \lim_{t\to\infty} \int d\bar{r} \int_{-\infty}^{+\infty} dz \int_{-\infty}^{t} \left( \bar{E}_z^{\mathrm{R}}(\bar{r},t') \cdot \frac{\partial}{\partial t'} \bar{D}_z^{\mathrm{R}}(\bar{r},t') + \bar{E}_\perp^{\mathrm{R}}(\bar{r},t') \cdot \frac{\partial}{\partial t'} \bar{D}_\perp^{\mathrm{R}}(\bar{r},t') + \bar{H}_z^{\mathrm{R}}(\bar{r},t') \cdot \frac{\partial}{\partial t'} \bar{B}_z^{\mathrm{R}}(\bar{r},t') + \bar{H}_\perp^{\mathrm{R}}(\bar{r},t') \cdot \frac{\partial}{\partial t'} \bar{B}_\perp^{\mathrm{R}}(\bar{r},t') \right) dt' \quad (15)$$

By substituting equations (13-14) into equation (15) and symmetrizing the result through the replacement of $\omega \leftrightarrows -\omega'$ and $\bar{\kappa}_\perp \leftrightarrows -\bar{\kappa}'_\perp$, we further have

$$W^s = \lim_{t\to\infty} 2\pi^2 \{\int_{-\infty}^{\infty} dz \int_{-\infty}^{\infty} d\omega d\omega' \int d\bar{\kappa}_\perp e^{-i(\omega-\omega')t} [\frac{\omega\varepsilon_z(z,\omega)-\omega'\varepsilon_z(z,-\omega')}{\omega-\omega'} E^R_{z,\omega,\bar{\kappa}_\perp}(z) E^R_{z,-\omega',-\bar{\kappa}_\perp}(z) +$$

$$\frac{\omega\varepsilon_\perp(z,\omega)-\omega'\varepsilon_\perp(z,-\omega')}{\omega-\omega'} \bar{E}^R_{\perp,\omega,\bar{\kappa}_\perp}(z) \cdot \bar{E}^R_{\perp,-\omega',-\bar{\kappa}_\perp}(z) + \frac{\omega\mu(z,\omega)-\omega'\mu(z,-\omega')}{\omega-\omega'} \bar{H}^R_{\perp,\omega,\bar{\kappa}_\perp}(z) \cdot \bar{H}^R_{\perp,-\omega',-\bar{\kappa}_\perp}(z)]\} \quad (16)$$

The fact of $\bar{H}^R_z = 0$ from equation (14) is used in equation (16). As shown in Fig. 1, we have the permittivity $\varepsilon_{z,\perp}(z) = \varepsilon_1(z<0) + \varepsilon_{gra,z,\perp}(z=0) + \varepsilon_2(z>0)$ in equation (16). If $\omega' \to \omega$, $\frac{\omega\varepsilon(z,\omega)-\omega'\varepsilon(z,-\omega')}{\omega-\omega'}$ is equivalent to $\frac{\partial\omega\varepsilon(\omega)}{\partial\omega}$, and $\frac{\omega\mu(z,\omega)-\omega'\mu(z,-\omega')}{\omega-\omega'}$ is equivalent to $\frac{\partial\omega\mu(\omega)}{\partial\omega}$. Note that in equation (16), the term with $\omega' \neq \omega$ would oscillate quickly with the time. Since $t \to \infty$, only the terms corresponding to $\omega' = \omega$ should be kept in the calculation of equation (16). On the other hand, the integration of $z$ in equation (16) would generate three parts of energy, which are below, within and above the monolayer graphene, respectively. Since $\varepsilon_{gra,\perp} = \frac{i\sigma_s}{\omega}\delta(z)$ and $\varepsilon_{gra,z}$ is a finite constant for the monolayer graphene, only the field components parallel to graphene plane would contribute to the calculation of the field energy inside the graphene.

Equation (16) can be simplified by using the knowledge of graphene plasmons. For transverse magnetic (TM) graphene plasmons, we have their dispersion as

$$\varepsilon_{1r}\sqrt{\varepsilon_{2r} - \frac{\kappa_\perp^2 c^2}{\omega^2}} + \varepsilon_{2r}\sqrt{\varepsilon_{1r} - \frac{\kappa_\perp^2 c^2}{\omega^2}} + \frac{\sigma_s}{c\varepsilon_0}\sqrt{\varepsilon_{1r} - \frac{\kappa_\perp^2 c^2}{\omega^2}}\sqrt{\varepsilon_{2r} - \frac{\kappa_\perp^2 c^2}{\omega^2}} = 0 \quad (17)$$

To facilitate the calculation, we can define a quantity $\bar{\zeta}_{\omega,\kappa_\perp}$ according to equations (17) as

$$\bar{\zeta}_{\omega,\kappa_\perp} = \varepsilon_{1r}\sqrt{\frac{\kappa_\perp^2 c^2}{\omega^2} - \varepsilon_{2r}} + \varepsilon_{2r}\sqrt{\frac{\kappa_\perp^2 c^2}{\omega^2} - \varepsilon_{1r}} + \frac{i\sigma_s}{c\varepsilon_0}\frac{\omega}{|\omega|}\sqrt{\frac{\kappa_\perp^2 c^2}{\omega^2} - \varepsilon_{1r}}\sqrt{\frac{\kappa_\perp^2 c^2}{\omega^2} - \varepsilon_{2r}} \quad (18)$$

As such, we can re-express the amplitudes $a_1$ and $a_2$ with the usage of equation (18) as

$$a_2 = -i\frac{\omega}{|\omega|}\frac{1}{\bar{\zeta}_{\omega,\kappa_\perp}}b_{2,\omega,\kappa_\perp}, \quad a_1 = -i\frac{\omega}{|\omega|}\frac{1}{\bar{\zeta}_{\omega,\kappa_\perp}}b_{1,\omega,\kappa_\perp} \quad (19)$$

Note that for the symmetric case discussed in this work, we set $\varepsilon_{1r} = \varepsilon_{2r} = \varepsilon_r$. Accordingly, we actually have $a_1 = -a_2 = a$, $\bar{\zeta}_{\omega,\kappa_\perp} = 2\varepsilon_r + \frac{i\sigma_s}{c\varepsilon_0}\frac{\omega}{|\omega|}\sqrt{\frac{\kappa_\perp^2 c^2}{\omega^2} - \varepsilon_r}$, and $b_{1,\omega,\kappa_\perp} = -b_{2,\omega,\kappa_\perp} = b_{\omega,\kappa_\perp} = \frac{\frac{v}{c}\frac{\kappa_\perp^2 c^2}{\omega^2 \varepsilon_r}\frac{\sigma_s}{c\varepsilon_0}}{(1-\frac{v^2}{c^2}\varepsilon_r+\frac{\kappa_\perp^2 v^2}{\omega^2})}$. This way, the plasmon energy in equation (16) can be expressed in terms of $b_{\omega,\kappa_\perp}$, that is,

$$W^{\mathrm{s}} = \lim_{t\to\infty} \frac{q^2}{32\pi^3\varepsilon_0} \int_0^\infty d\kappa_\perp^2 \int_{-\infty}^\infty d\omega \int_{-\infty}^\infty d\omega' \frac{e^{-i(\omega-\omega')t}}{\bar{\zeta}_{\omega,\kappa_\perp}\bar{\zeta}_{-\omega',\kappa_\perp}} \frac{|b_{\omega,\kappa_\perp}|^2}{\omega^2} \left\{ \frac{\frac{\partial[\omega\varepsilon_{\mathrm{r}}(\omega)]}{\partial\omega}\left(2-\frac{\varepsilon_{\mathrm{r}}(\omega)\omega^2}{\kappa_\perp^2 c^2}\right)+\frac{\omega^2\varepsilon_{\mathrm{r}}^2(\omega)}{\kappa_\perp^2 c^2}}{\sqrt{\kappa_\perp^2 - \frac{\omega^2}{c^2}\varepsilon_{\mathrm{r}}(\omega)}} + \frac{\partial\left[\frac{i\sigma_{\mathrm{s}}(\omega)}{\varepsilon_0}\right]}{\partial\omega} \frac{\omega^2}{\kappa_\perp^2 c^2} \left(\frac{\kappa_\perp^2 c^2}{\omega^2} - \varepsilon_{\mathrm{r}}(\omega)\right) \right\} \tag{20}$$

We highlight that the term in the bracket of equation (20) is an even function about $\omega$ and $\kappa_\perp$. If $f(\omega)$ is an even function, that is, $f(\omega) = f(-\omega)$, we have

$$\lim_{t\to\infty} \int_{-\infty}^\infty d\omega f(\omega) \int_{-\infty}^\infty d\omega' \frac{e^{-i(\omega-\omega')t}}{\bar{\zeta}_{\omega,\kappa_\perp}\bar{\zeta}_{-\omega',\kappa_\perp}} = 8\pi^2 \int_0^\infty d\omega f(\omega) \frac{\delta(\bar{\zeta}_{\omega,\kappa_\perp})}{(\partial/\partial\omega)\bar{\zeta}_{\omega,\kappa_\perp}} \tag{21}$$

From equation (21), we can further simplify equation (20) to the following

$$W^{\mathrm{s}} = \int_0^\infty w^{\mathrm{s}}(\omega)d\omega = \frac{q^2}{4\pi\varepsilon_0} \int_0^\infty \frac{d\omega}{\omega^2} \int_0^\infty d\kappa_\perp^2 \frac{|b_{\omega,\kappa_\perp}|^2 \delta(\bar{\zeta}_{\omega,\kappa_\perp})}{(\partial/\partial\omega)\bar{\zeta}_{\omega,\kappa_\perp}} \frac{\frac{\partial[\omega\varepsilon_{\mathrm{r}}(\omega)]}{\partial\omega}\left(2-\frac{\varepsilon_{\mathrm{r}}(\omega)\omega^2}{\kappa_\perp^2 c^2}\right)+\frac{\omega^2\varepsilon_{\mathrm{r}}^2(\omega)}{\kappa_\perp^2 c^2}}{\sqrt{\kappa_\perp^2 - \frac{\omega^2}{c^2}\varepsilon_{\mathrm{r}}(\omega)}} \tag{22}$$

where $w^{\mathrm{s}}(\omega)$ is the spectrum of the excited graphene plasmons during the process of transition radiation.

After the integration of $d\kappa_\perp^2$, we can finally obtain the energy spectrum of graphene plasmons as

$$w^{\mathrm{s}}(\omega) = \frac{q^2\varepsilon_{\mathrm{r}}}{\pi\varepsilon_0 c} \frac{\frac{v^2}{c^2}|\frac{\sigma_{\mathrm{s}}}{\varepsilon_0 c}-4\varepsilon_{\mathrm{r}}\frac{\varepsilon_0 c}{\sigma_{\mathrm{s}}}|^2}{[1-\frac{v^2}{c^2}(\frac{2\varepsilon_{\mathrm{r}}\varepsilon_0 c}{\sigma_{\mathrm{s}}})^2]^2} \cdot \frac{1}{-\frac{i\sigma_{\mathrm{s}}}{\varepsilon_0 c}[2\varepsilon_{\mathrm{r}}-\frac{1}{2}(\frac{\sigma_{\mathrm{s}}}{\varepsilon_0 c})^2]} \tag{23}$$

Figure 3 shows the energy spectrum of excited graphene plasmons according to equation (23). We first discuss the influence of the particle velocity on the excitation of graphene plasmons in Fig. 3a. From Fig. 3a, a lower particle velocity can lead to the more efficient excitation of graphene plasmons in a broader frequency range. This is mainly because the particle with a smaller velocity can have a longer time of interaction with the monolayer graphene. This way, more energy can be extracted from the swift electron and transformed into the graphene plasmons. Moreover, the peak-intensity frequency for the excited graphene plasmons increases if the particle velocity decreases. As such, it is reasonable to conclude that the more excitation of graphene plasmons favours a lower particle velocity.

On the other hand, we discuss the influence of the chemical potential of graphene on the excitation of graphene plasmons at a fixed particle velocity in Fig. 3b. From Fig. 3b, a higher chemical potential leads to the more efficient excitation of graphene plasmons at higher frequencies. Meanwhile, the peak-intensity frequency for the excited graphene plasmons also increases with the chemical potential.

**Far-field analysis of the transition radiation from graphene**

Figure 4 shows the spatial distribution of excited waves in the far field when a swift electron perpendicularly crosses a graphene monolayer. To be specific, Fig. 4 studies the influence of the particle velocity on the transition radiation from graphene in the far field. From Fig. 4, when we increase the particle velocity, the intensity of excited photons becomes stronger, along with a better radiation directivity. Especially, when $v \to c$ [Fig. 4f], most of the excited photons would propagate along a direction almost perpendicular to the graphene plane. The underlying reason is that for a larger particle velocity, a more spatially-extended charge disturbance or polarization current is induced in the graphene plane in the neighborhood of the charge trajectory.

To quantitatively discuss the far-field radiation, below we analytically calculate the backward radiation energy $W_1$ of excited photons, namely the energy of photons emitted into medium 1. In principle, the backward radiation energy can be obtained by integrating the field energy density over all space in medium 1 at $t \to \infty$. At $t \to \infty$, the emitted radiation field, which is a photonic pulse in the time domain, is already far away from the boundary and well separated from the charge field. Moreover, if we move the coordinate origin along the particle trajectory into the position having the emitted photonic pulse, the integration with respect to *z* can be performed from $-\infty$ to $+\infty$, since the field of excited photons is attenuated in both directions away from the central position of the photonic pulse.

For the excited photons, the electric and magnetic energy densities are equal in free space. This way, the backward radiation energy can be readily expressed as

$$W_1 = \int dxdy \int_{-\infty}^{+\infty} dz \cdot \varepsilon_1 \left| \bar{E}_1^R(\bar{r},t) \right|^2 \quad (24)$$

$$\left| \bar{E}_1^R(\bar{r},t) \right|^2 = \int \bar{E}_{1|\bar{\kappa}_\perp,\omega}^R(\bar{r},t) \cdot \bar{E}_{1|\bar{\kappa}'_\perp,\omega'}^{R\;*}(z) e^{i[(\bar{\kappa}_\perp-\bar{\kappa}'_\perp)\cdot\bar{r}_\perp-(\omega-\omega')t]} \, d\bar{\kappa}_\perp d\bar{\kappa}'_\perp d\omega d\omega' \quad (25)$$

By substitute equation (25) into (24) and performing the integration over $d\bar{\kappa}'_\perp$ and $d\omega'$, we obtain

$$W_1 = 2\int_0^{+\infty} d\omega \int \varepsilon_1 |a_1|^2 \left(\frac{q}{\omega\varepsilon_0(2\pi)^3}\right)^2 \frac{\omega^2}{c\kappa_\perp^2} \sqrt{\varepsilon_{1r}} \sqrt{1 - \frac{\kappa_\perp^2 c^2}{\omega^2 \varepsilon_{1r}}} (2\pi)^3 d\bar{\kappa}_\perp \quad (26)$$

For the emitted photons in medium 1, we have $\kappa_\perp^2 < \frac{\omega^2}{c^2}\varepsilon_{1r}$. As such, the integration over $d\bar{\kappa}_\perp$ in equation (26) is operated in the range $\kappa_\perp^2 < \frac{\omega^2}{c^2}\varepsilon_{1r}$. We can further express $\kappa_\perp = \frac{\omega}{c}\sqrt{\varepsilon_{1r}}\sin\theta$, by defining $\theta$ to be the angle between the wavevector of excite photons and $-\bar{v}$; see the schematic illustration in Fig. 4a. Then we have $2\pi\kappa_\perp d\kappa_\perp = 2\pi(\frac{\omega^2}{c^2}\varepsilon_{1r})\sin\theta\cos\theta d\theta$. By substituting this relation into equation (26), we obtain

$$W_1 = \int_0^{+\infty}\int_0^{\pi/2} U_1(\omega,\theta,\beta)\cdot(2\pi\sin\theta)d\theta d\omega = \int_0^{+\infty}\int_0^{\pi/2} \frac{\varepsilon_{1r}^{3/2}q^2\cos^2\theta}{4\pi^3\varepsilon_0 c\sin^2\theta}|a_1|^2\cdot(2\pi\sin\theta)d\theta d\omega \qquad (27)$$

Here $U_1(\omega,\theta,\beta) = \frac{\varepsilon_{1r}^{3/2}q^2\cos^2\theta}{4\pi^3\varepsilon_0 c\sin^2\theta}|a_1|^2$ is the angular spectral energy density of the backward radiation. By substituting the expression for $a_1$ in equation (9), we can write the backward angular spectral energy density explicitly as

$$U_1(\omega,\theta,\beta) = \frac{\sqrt{\varepsilon_r}q^2\beta^2\cos^2\theta\sin^2\theta}{4\pi^3\varepsilon_0 c(1-\varepsilon_r\beta^2\cos^2\theta)^2}\left|\frac{\frac{\sigma_s}{c\varepsilon_0}}{2\sqrt{\varepsilon_r}+\frac{\sigma_s}{c\varepsilon_0}\cos\theta}\right|^2 \qquad (28)$$

In addition, if we define $W_1 = \int_0^\infty w_1(\omega)d\omega$, the energy spectrum of backward radiation can be derived as

$$w_1(\omega,\beta) = \int_0^{\pi/2} U_1(\omega,\theta,\beta)\cdot(2\pi\sin\theta)d\theta \qquad (29)$$

The calculation of the energy spectrum for the forward radiation $w_2(\omega)$ can be performed by following a similar calculation procedure of $w_1(\omega)$. Then the total spectrum of emitted photon can be written as $w(\omega) = w_1(\omega) + w_2(\omega)$. Since $\varepsilon_{1r} = \varepsilon_{2r} = \varepsilon_r$ in this work, we have $w_1(\omega) = w_2(\omega)$ and $w(\omega) = 2w_1(\omega)$.

Figure 5a shows the backward angular spectral energy density of excited photons as a function of the radiation angle $\theta$ and the particle velocity at a fixed frequency of 10 THz, according to equation (28). From Fig. 5a, the energy of emitted photons increases with the particle velocity. Meanwhile, we find that the angular spectral energy density has a smaller peak-intensity angle for a larger electron velocity [Fig. 5a]. Particularly, under the condition of $\varepsilon_r = 1$, the peak-intensity angle appears approximately at $\theta = \frac{\sqrt{1-\beta^2}}{\sqrt{2}\beta}$ if $\beta \to 1$, according to equation (28). These results are in accordance with the far-field distribution of excited waves in Fig. 4.

Figure 5b shows the spectrum of excited photons. From Fig. 5b, a faster electron can give rise to the more efficient excitation of photons in a wider frequency range. In other words, the excitation of photons prefers a larger particle velocity, which is distinct from the excitation of graphene plasmons in Fig. 3. Last but not least, we highlight that from the comparison between Figs. 3&5, the energy of excited graphene plasmons is around two orders of magnitude larger than that of emitted photons during the process of transition radiation. This way, most of the electron energy loss is converted into the graphene plasmons during the transition radiation process.

**Conclusion**

In conclusion, we have systematically investigated the transition radiation from a monolayer graphene in the frequency domain. From the spatial distribution of the radiation field, both the graphene plasmons and photons can be excited in a broad bandwidth. From the analytical spectra of the excited photons and plasmons, we find that the excitation of graphene plasmons prefers a lower particle velocity, while the excitation of photons favours a higher particle velocity. Our results might provide theoretical guidance for the design of novel on-chip radiation sources [50] based on graphene. On the other hand, we highlight that graphene is just a typical example of 2D materials. Due to the abundance of 2D materials and emerging physics (e.g., twisted optics) in van der Waals heterostructures [29, 30], our work may ignite more fundamental researches about the electromagnetic radiation from the interaction between swift charged particles and various van der Waals materials or heterostructures, such as the in-plane anisotropic black phosphorous (BP) [35, 36], the uniaxial and hyperbolic hexagonal boron nitride (*h*BN) [29, 30], and the biaxial and hyperbolic α-phase molybdenum trioxide (α-$MoO_3$) [39-43], and twisted atomic bilayers [44, 45].

**References**
[1] V. L. Ginzburg, V. N. Tsytovich, Several problems of the theory of transition radiation and transition scattering. Phys. Rep. **49**, 1 (1979).
[2] V. L. Ginzburg, V. N. Tsytovich, *Transition Radiation and Transition Scattering* (Adam Higler, 1990).
[3] I. Frank, I. Tamm, Coherent visible radiation from fast electrons passing through matter. Comptes Rendus (Dokl.) Acad. Sci. USSR **14**, 109 (1937).
[4] Z. Su, B. Xiong, Y. Xu, Z. Cai, J. Yin, R. Peng, Y. Liu, Manipulating Cherenkov radiation and Smith-Purcell radiation by artificial structures. Adv. Optical Mater. **7**, 1801666 (2019).
[5] F. J. G. Abajo, Optical excitations in electron microscopy. Rev. Mod. Phys. **82**, 209 (2010).
[6] H. Hu, X. Lin, J. Zhang, D. Liu, P. Genevet, B. Zhang, and Y. Luo, Nonlocality induced Cherenkov threshold. Laser & Photonics Reviews 2020, 202000149 (2020). [doi.org/10.1002/lpor.202000149]
[7] F. Liu, L. Xiao, Y. Ye, M. Wang, K. Cui, X. Feng, W. Zhang, Y. Huang, Integrated Cherenkov radiation emitter eliminating the electron velocity threshold. Nature Photonics **11**, 289 (2017).
[8] S. Casalbuoni, B. Schmidt, P. Schmüser, V. Arsov, and S. Wesch, Ultrabroadband terahertz source and beamline based on coherent transition radiation. Physical Review Special Topics-Accelerators and Beams **12**, 030705 (2009).

[9] G. Liao, Y. Li, Y. Zhang, H. Liu, X. Ge, S. Yang, W. Wei, X. Yuan, Y. Deng, B. Zhu, Z. Zhang, W. Wang, Z. Sheng, L. Chen, X. Lu, J. Ma, X. Wang, J. Zhang, Demonstration of coherent terahertz transition radiation from relativistic laser-solid interactions. Phys. Rev. Lett. **116**, 205003 (2016).
[10] M. A. Piestrup, D. G. Boyers, C. I. Pincus, Qiang Li, G. D. Hallewell, M. J. Moran, D. M. Skopik, R. M. Silzer, X. K. Maruyama, D. D. Snyder, and G. B. Rothbart, Observation of soft-x-ray spatial coherence from resonance transition radiation. Phys. Rev. A **45**, 1183 (1992).
[11] X. Lin, S. Easo, Y. Shen, H. Chen, B. Zhang, J. D. Joannopoulos, M. Soljačić, I. Kaminer, Controlling Cherenkov angles with resonance transition radiation. Nature Physics **14**, 816 (2018).
[12] A. N. Chu, M. A. Piestrup, T. W. Barbee Jr, R. H. Pantell, Transition radiation as a source of X rays. Journal of Applied Physics **51**, 1290 (1980).
[13] S. Paklue, S. Rimjaem, J. Saisut, C. Thongbai, Coherent THz transition radiation for polarization imaging experiments. Nuclear Inst. and Methods in Physics Research B **464**, 28 (2020).
[14] A. E. Kaplan, C. T. Law, and P. L. Shkolnikov, X-ray narrow-line transition radiation source based on low-energy electron beams traversing a multilayer nanostructure. Phys. Rev. E **52**, 6795 (1995).
[15] J. D. Jackson, *Classical Electrodynamics* (John Wiley & Sons, 1999).
[16] B. Dolgoshein, Transition radiation detectors. Nuclear Instruments and Methods in Physics Research A **326**, 434 (1993).
[17] X. Artru, G. B. Yodh, G. Mennessier, Practical theory of the multilayered transition radiation detector. Phys. Rev. D **12**, 1289 (1975).
[18] K. D. Vries, S. Prohira, Coherent transition radiation from the geomagnetically induced current in cosmic-ray air showers: Implications for the Anomalous Events Observed by ANITA. Phys. Rev. Lett. **123**, 091102 (2019).
[19] G. Adams, V. Burkert, R. Carl, T. Carstens, V. Frolov, L. Houghtlin, G. Jacobs, M. Kossov, M. Klusman, B. Kross, M. Onuk, J. Napolitano, J.W. Price, C. Riggs, Y. Sharabian, A. Stavinsky, L.C. Smith, W.A. Stephens, P. Stoler, W. Tuzel, K. Ullrich, A. Vlassov, A. Weisenberger, M. Witkowski, B. Wojtekhowski, P.F. Yergin, C. Zorn, The CLAS Cherenkov detector. Nuclear Instruments and Methods in Physics Research A **465**, 414 (2001).
[20] W. C. Haxton, Nuclear response of water Cherenkov detectors to supernova and solar neutrinos. Phys. Rev. D **36**, 2283 (1987).
[21] F. J. G. Abajo, Multiple excitation of confined graphene plasmons by single free electrons. ACS Nano **7**, 11409 (2013).
[22] Z. L. Mišković, S. Segui, J. L. Gervasoni, N. R. Arista, Energy losses and transition radiation produced by the interaction of charged particles with a graphene sheet. Phys. Rev. B **94**, 125414 (2016).
[23] K. Akbari, Z. L. Mišković, S. Segui, J. L. Gervasoni, N. R. Arista, Energy losses and transition radiation in multilayer graphene traversed by a fast charged particle. ACS Photonics **4**, 1980 (2017).
[24] X. Lin, I. Kaminer, X. Shi, F. Gao, Z. Yang, Z. Gao, H. Buljan, J. D. Joannopoulos, M. Soljačić, H. Chen, B. Zhang, Splashing transients of 2D plasmons launched by swift electrons. Science Advances **3**, e1601192 (2017).
[25] T. Ochiai, Efficiency and angular distribution of graphene-plasmon excitation by electron beam. Journal of the Physical Society of Japan **83**, 054705 (2014).
[26] K. Zhang, X. Chen, C. Sheng, K. J. A. Ooi, L. K. Ang, X. Yuan, Transition radiation from graphene plasmons by a bunch beam in the terahertz regime. Optics Letters **25**, 20477 (2017).
[27] S. Gong, M. Hu, Z. Wu, H. Pan, H. Wang, K. Zhang, R. Zhong, J. Zhou, T. Zhao, D. Liu, W. Wang, C. Zhang, S. Liu, Direction controllable inverse transition radiation from the spatial dispersion in a graphene-dielectric stack. Photonics Research **7**, 1154 (2019).
[28] M. H. Gass, U. Bangert, A. L. Bleloch, P. Wang, R. R. Nair, A. K. Geim, Free-standing graphene at atomic resolution. Nature Nanotechnology **3**, 676 (2008).
[29] T. Low, A. Chaves, J. D. Caldwell, A. Kumar, N. X. Fang, P. Avouris, T. F. Heinz, F. Guinea, L. Martin-Moreno, F. Koppens, Polaritons in layered two-dimensional materials. Nature Materials. **16**, 182 (2017).
[30] D. N. Basov, M. M. Fogler, F. J. García de Abajo, Polaritons in van der Waals materials. Science **354**, 1992 (2016).
[31] X. Lin, Y. Yang, N. Rivera, J. J. López, Y. Shen, I. Kaminer, H. Chen, B. Zhang, J. D. Joannopoulos, M. Soljačić, All-angle negative refraction of highly squeezed plasmon and phonon polaritons in graphene-boron

nitride heterostructures. Proceedings of the National Academy of Sciences of the United States of America **114**, 6717 (2017).

[32] X. Zhang, H. Hu, X. Lin, L. Shen, B. Zhang, H. Chen, Confined transverse-electric graphene plasmons in negative refractive-index systems. npj 2D Materials and Applications **4**, 25 (2020).

[33] C. Wang, C. Qian, H. Hu, L. Shen, Z. Wang, H. Wang, Z. Xu, B. Zhang, H. Chen, X. Lin, Superscattering of light in refractive-index near-zero Environments. Progress In Electromagnetics Research **168**, 15 (2020).

[34] X. Lin, Y. Shen, I. Kaminer, H. Chen, M. Soljačić, Transverse-electric Brewster effect enabled by nonmagnetic two-dimensional materials. Phys. Rev. A **94**, 023836 (2016).

[35] T. Low, R. Roldán, H. Wang, F. Xia, P. Avouris, L. M. Moreno, F. Guinea, Plasmons and screening in monolayer and multilayer black phosphorus. Phys. Rev. Lett. **113**, 106802 (2014).

[36] M. Renuka, X. Lin, Z. Wang, L. Shen, B. Zheng, H. Wang, and H. Chen, Dispersion engineering of hyperbolic plasmons in bilayer 2D materials. Optics Letters **43**, 5737 (2018).

[37] S. Shah, X. Lin, L. Shen, M. Renuka, B. Zhang, and H. Chen, Interferenceless polarization splitting through nanoscale van der Waals heterostructures. Phys. Rev. Applied **10**, 034025 (2018).

[38] M. Musa, M. Renuka, X. Lin, R. Li, H. Wang, E. Li, B. Zhang, and H. Chen, Confined transverse electric phonon polaritons in hexagonal boron nitrides. 2D Materials **5**, 015018 (2018).

[39] G. Hu, Q. Ou, G. Si, Y. Wu, J. Wu, Z. Dai, A. Krasnok, Y. Mazor, Q. Zhang, Q. Bao, C. Qiu, A. Alù, Topological polaritons and photonic magic angles in twisted α-MoO3 bilayers. Nature **582**, 209 (2020).

[40] M. Chen, X. Lin, T. H Dinh, Z. Zheng, J. Shen, Q. Ma, H. Chen, P. J. Herrero, S. Dai, Configurable phonon polaritons in twisted α-MoO$_3$. Nature Materials, in press (2020). [doi.org/10.1038/s41563-020-0732-6]

[41] J. Duan, N. C. Robayna, J. T. Gutiérrez, G. Álvarez-Pérez, I. Prieto, J. M. Sánchez, A. Y. Nikitin, P. A. González, Twisted nano-optics: manipulating light at the nanoscale with twisted phonon polaritonic slabs. Nano Letter **20**, 5323 (2020).

[42] Z. Zheng, F. Sun, W. Huang, J. Jiang, R. Zhan, Y. Ke, H. Chen, S. Deng, Phonon polaritons in twisted double-layers of hyperbolic van der Waals crystals. Nano Letter **20**, 5301 (2020).

[43] W. Ma, P. A. González, S. Li, A. Y. Nikitin, J. Yuan, J. M. Sánchez, J. T. Gutiérrez, I. Amenabar, P. Li, S. Vélez, C. Tollan, Z. Dai, Y. Zhang, S. Sriram, K. K. Zadeh, S.T. Lee, R. Hillenbrand, Q. Bao, In-plane anisotropic and ultra-low-loss polaritons in a natural van der Waals crystal. Nature **562**, 557 (2018).

[44] T. Stauber, T. Low, G. Gómez-Santos, Chiral response of twisted bilayer graphene. Phys. Rev. Lett. **120**, 046801 (2018).

[45] X. Lin, Z. Liu, T. Stauber, G. G. Santos, F. Gao, H. Chen, B. Zhang, T. Low, Chiral plasmons with twisted atomic bilayers. Phys. Rev. Lett. **125**, 077401 (2020).

[46] X. Shi, X. Lin, I. Kaminer, F. Gao, Z. Yang, J. D. Joannopoulos, M. Soljačić, B. Zhang, Superlight inverse Doppler effect. Nature Physics **14**, 1001 (2018).

[47] X. Lin, B. Zhang, Normal Doppler frequency shift in negative refractive-index systems. Laser & Photonics Reviews **13**, 1900081 (2019).

[48] Y. Jiang, X. Lin, T. Low, B. Zhang, H. Chen, Group-velocity-controlled and gate-tunable directional excitation of polaritons in graphene-boron nitride heterostructures. Laser & Photonics Reviews **12**, 1800049 (2018).

[49] X. Lin, R. Li, F. Gao, E. Li, X. Zhang, B. Zhang, H. Chen, Loss induced amplification of graphene plasmons. Optics Letters **41**, 681 (2016).

[50] L. Shen, X, Lin, M. Y. Shalaginov, T. Low, X. Zhang, B. Zhang, H. Chen, Broadband enhancement of on-chip single-photon extraction via tilted hyperbolic metamaterials featured. Applied Physics Reviews **7**, 021403 (2020).

**Acknowledgments**

The work at Zhejiang University was sponsored by the National Natural Science Foundation of China (NNSFC) under Grants No. 61625502, No.11961141010, and No. 61975176, the Top-Notch Young Talents Program of China, and the Fundamental Research Funds for the Central Universities.

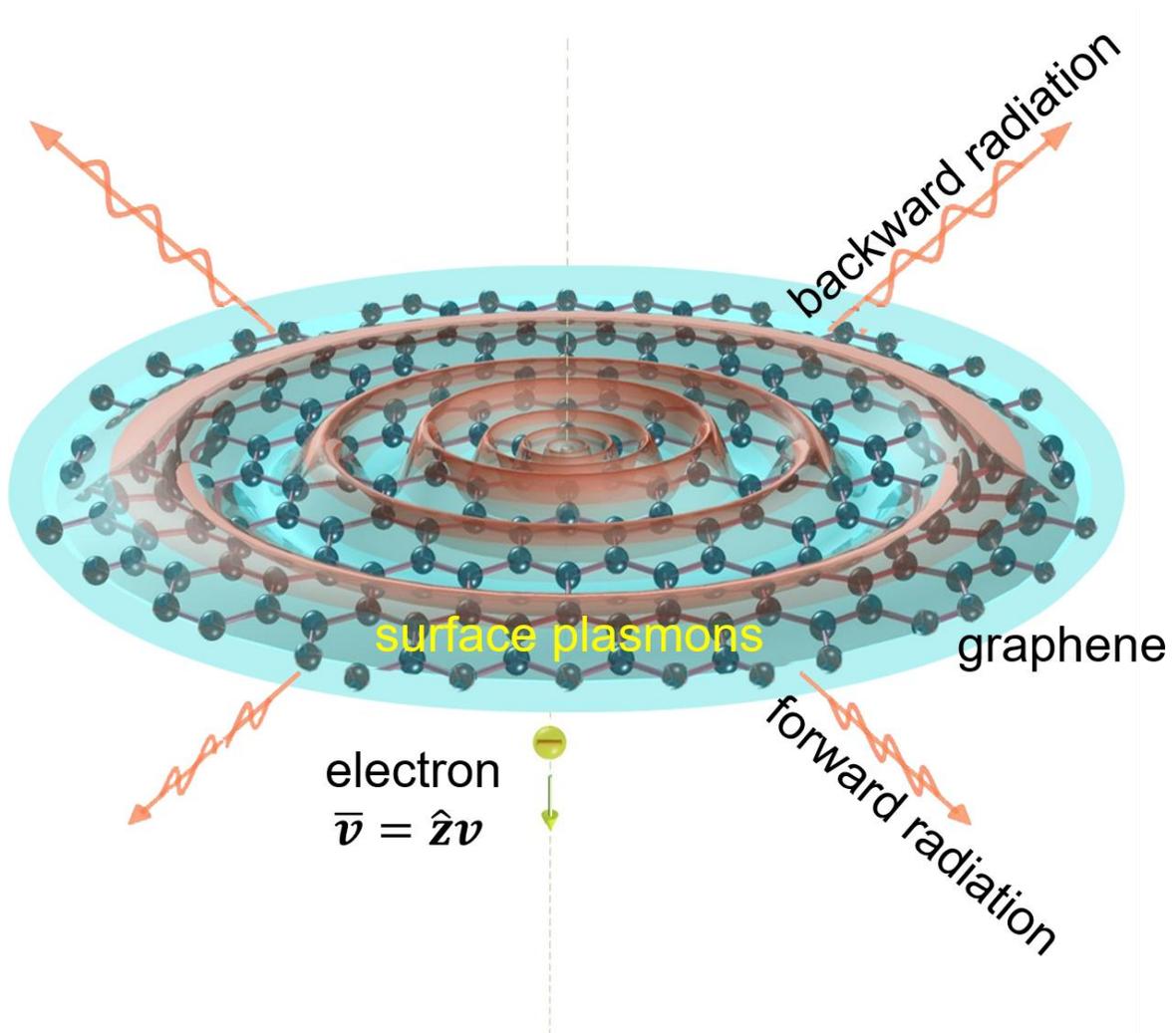

**Figure 1 | Schematic of the transition radiation from a monolayer graphene.** After the perpendicular penetration of swift electrons through the graphene layer, both graphene plasmons and freely propagating light are excited. Without specific specification, we set the chemical potential of graphene to be $\mu_c = 0.4$ eV.

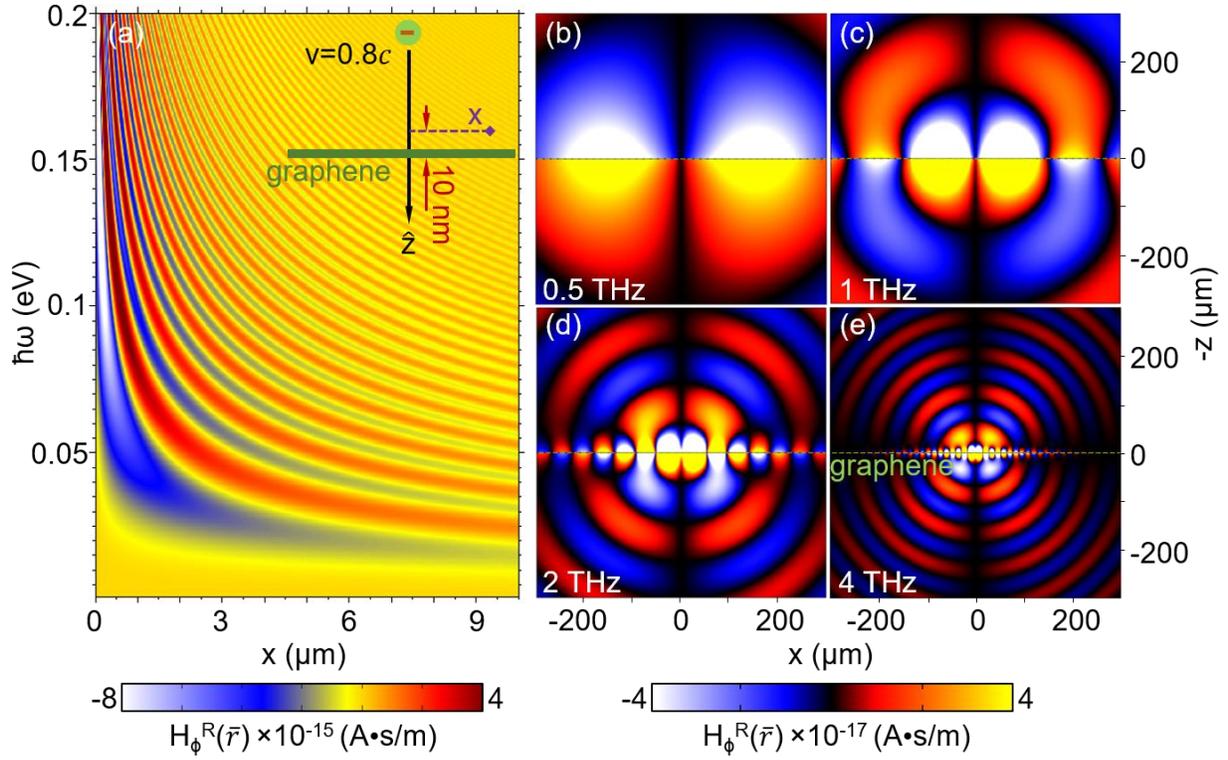

**Figure 2 | Near-field distribution of excited waves when a swift electron perpendicularly penetrates through a monolayer graphene.** (a) Distribution of the excited graphene plasmons. Here we plot the magnetic field $H_\phi^R(\bar{r})$ as a function of frequency along a line, which is parallel and close to the graphene plane with a vertical distance of 10 nm; see the schematic illustration in the inset. (b-e) Near-field distribution of excited waves at (b) 0.5 THz, (c) 1 THz, (d) 2 THz and (e) 4 THz. Here, $v = 0.8c$.

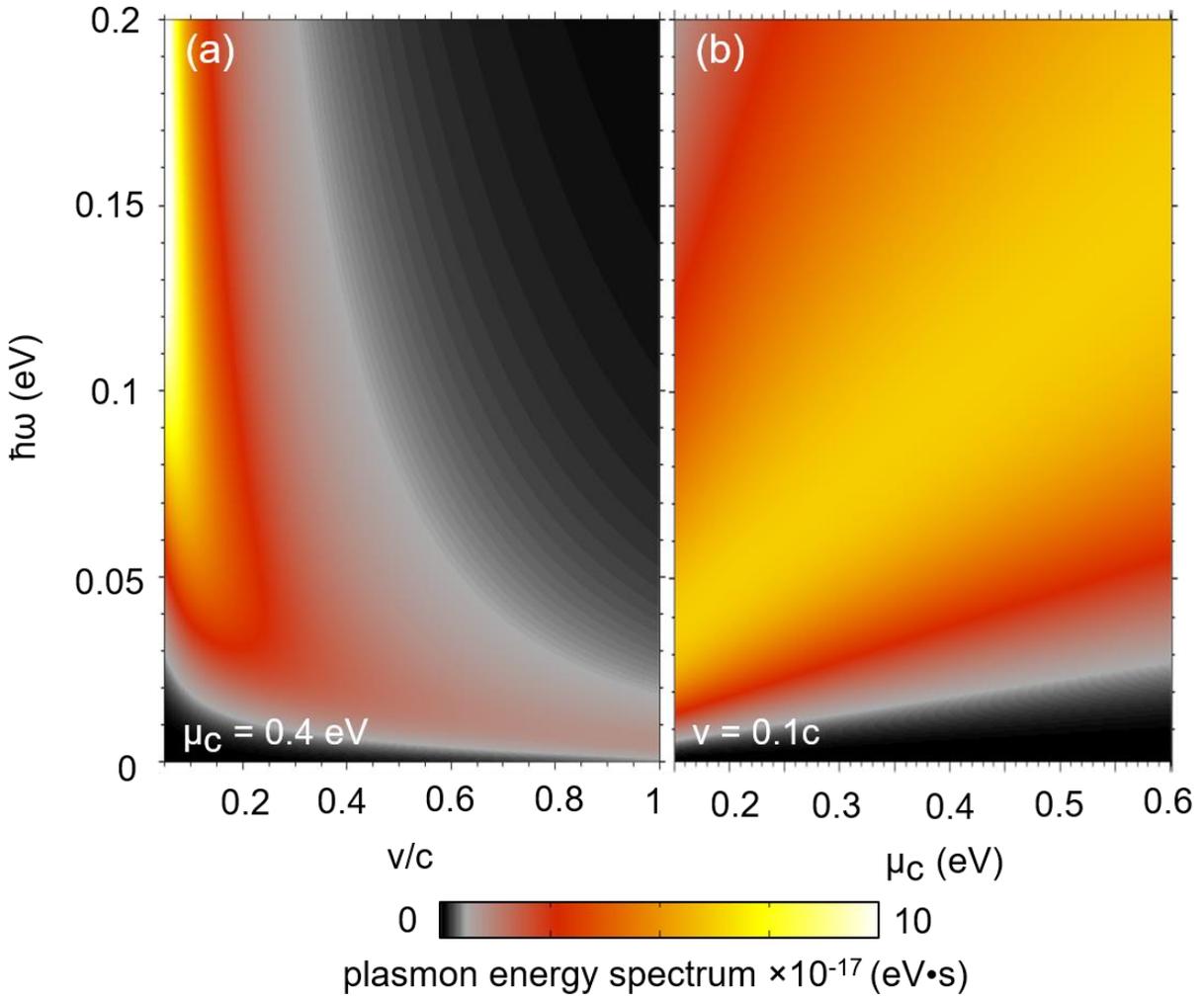

**Figure 3 | Spectrum of the excited graphene plasmons.** The structural setup is the same as Fig. 1. (a) Influence of the electron velocity $v$ on the excitation of graphene plasmons. We have $\mu_c = 0.4$ eV in (a). (b) Influence of $\mu_c$ on the excitation of graphene plasmons. We set $v = 0.1c$ in (b).

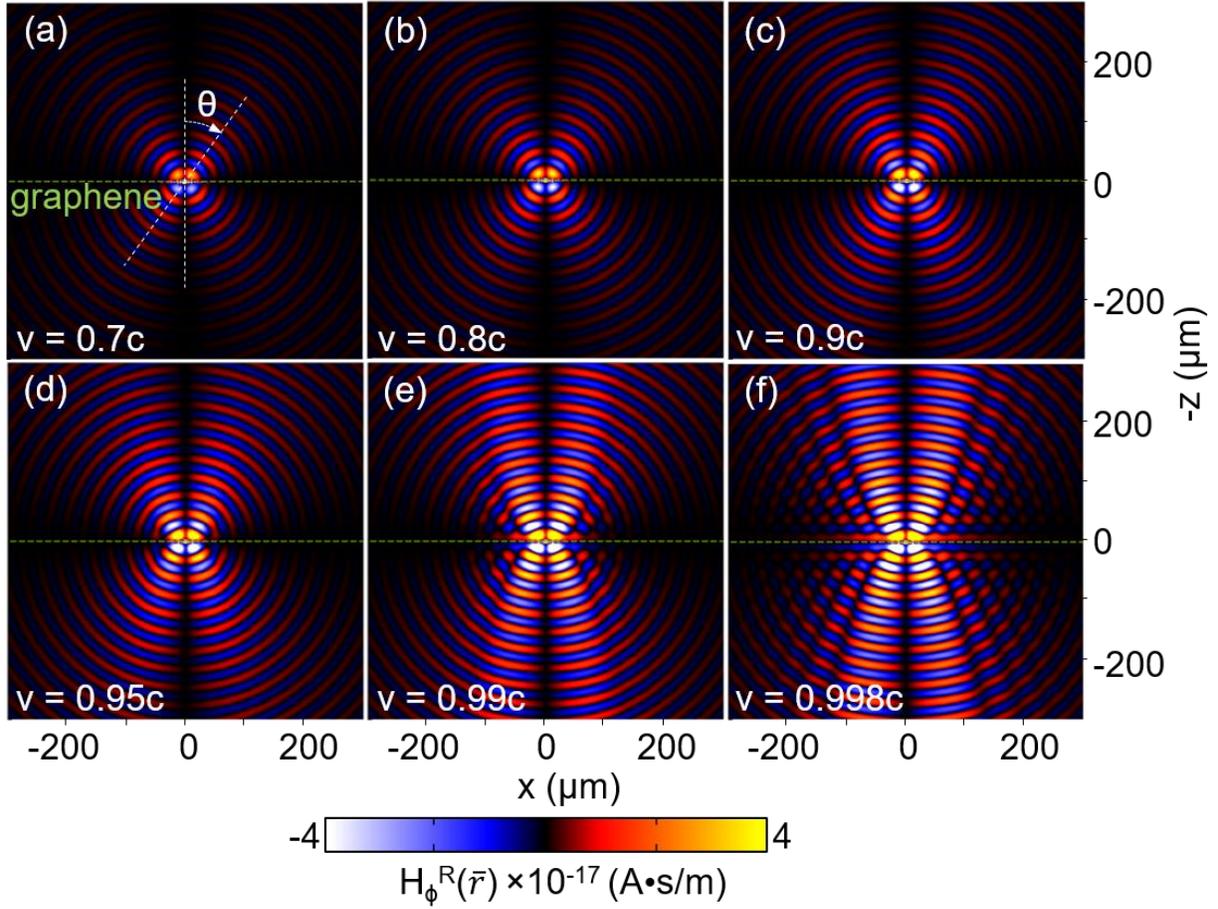

**Figure 4 | Far-field distribution of the excited waves when a swift electron perpendicularly crosses a monolayer graphene.** (a-f) Spatial Distribution of the magnetic field $H_\phi^R(\bar{r})$ with (a) $v = 0.7c$, (b) $v = 0.8c$, (c) $v = 0.9c$, (d) $v = 0.95c$, (e) $v = 0.99c$ and (f) $v = 0.998c$. Here the working frequency is 10 THz.

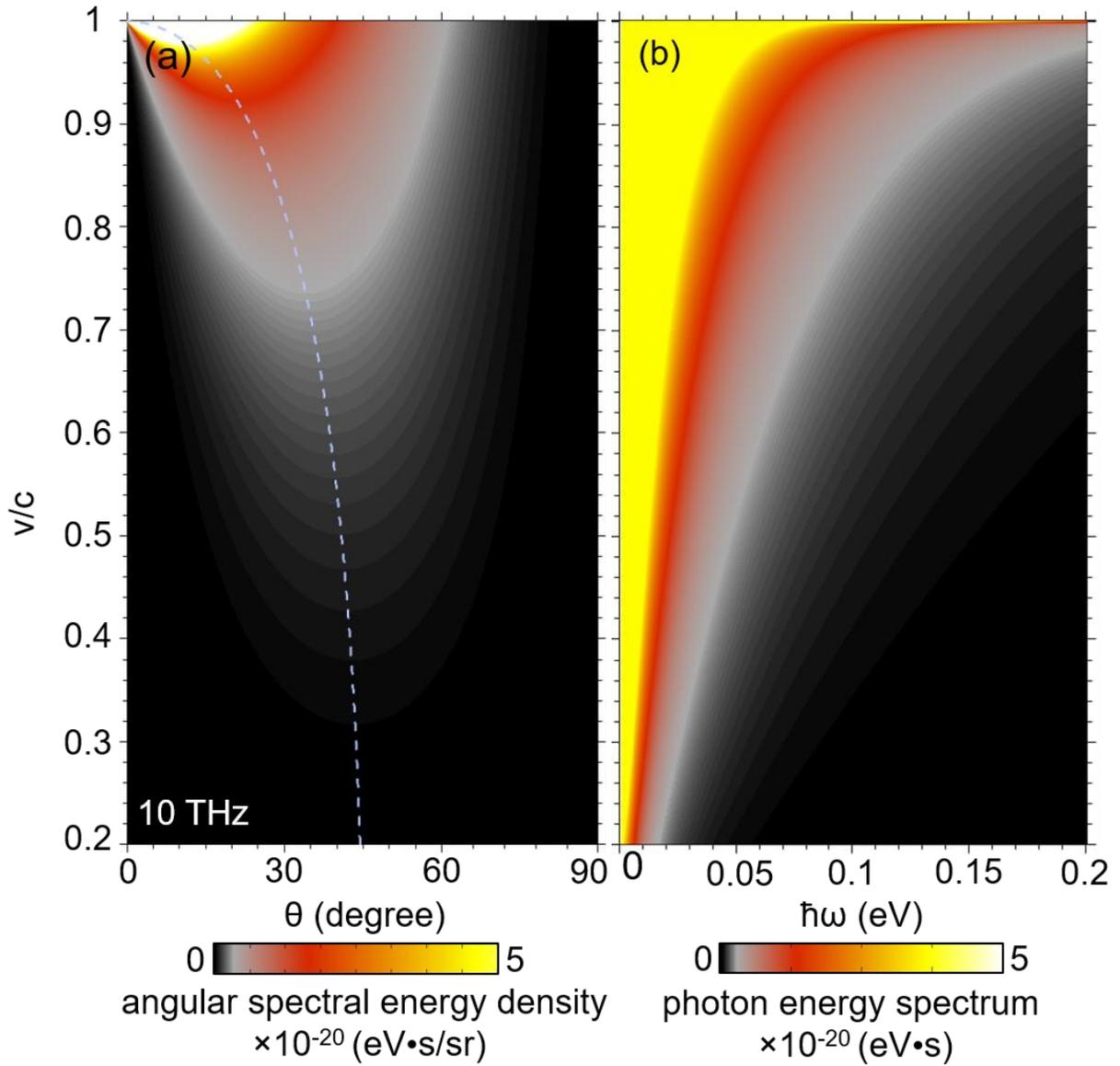

**Figure 5 | Spectrum of the excited photons.** The structural setup is the same as Fig. 1. (a) Angular spectral energy density of the backward radiation as a function of the radiation angle $\theta$ and the electron velocity $v$ at 10 THz. The peak-intensity angle is highlighted by the dashed line, and it approaches to zero if the value of $v$ approaches to $c$. (b) Spectrum of the excited photons as a function of the frequency and $v$.